\documentclass[10pt,twocolumn,letterpaper]{article}

\usepackage{3dv}

\makeatletter
\@namedef{ver@everyshi.sty}{}
\makeatother

\usepackage{amsmath}
\usepackage{amssymb}
\usepackage{enumitem} 													
\usepackage{algorithm}
\usepackage[noend]{algpseudocode}
\usepackage{tikz}
\usepackage{tikzscale}
\usetikzlibrary{arrows}
\usepackage{pgfplots}
\pgfplotsset{compat=1.12}
\usepackage{subfig}
\usepackage{wrapfig}
\usepackage{color,soul}
\soulregister\cite7
\soulregister\ref7
\soulregister\pageref7
\usepackage{booktabs}

\DeclareMathOperator{\arclen}{chordlen}

\def\*#1{\mathbf{#1}}

\renewcommand{\epsilon}{\varepsilon}
\renewcommand{\hat}{\widehat}

\newtheorem{remark}{Remark}

\threedvfinalcopy 

\def\threedvPaperID{109} 

\begin{document}

\title{Deep Learning Parametrization for B-Spline Curve Approximation}

\author{Pascal Laube*
\and
Matthias O. Franz*
\and
Georg Umlauf*
\and
*\textit{Institute for Optical Systems, University of Applied Sciences Konstanz, Germany}\\
{\tt\small pascal.laube@gmail.com}
}

\maketitle
 \thispagestyle{empty}

\begin{abstract}
In this paper we present a method using deep learning to compute parametrizations for B-spline curve approximation. Existing methods consider the computation of parametric values and a knot vector as separate problems. We propose to train interdependent deep neural networks to predict parametric values and knots. We show that it is possible to include B-spline curve approximation directly into the neural network architecture. The resulting parametrizations yield tight approximations and are able to outperform state-of-the-art methods.
\end{abstract}

\section{Introduction}
In the parametrization of B-spline curve approximation, both parametric values
and a suitable knot vector have to be computed.
Since the approximation quality strongly depends on the
parametrization this is a key problem for many applications.
However, finding good parametrizations is a complex task and computationally difficult.
A good parametrization may be defined as a set of parametric values relating to
data points and a knot vector that leads to a minimal deviation between data points
and the approximating curve.
Typically, an error threshold needs to be satisfied. In addition, the approximation should result in as few control points as possible. Thus, the number of knots has
to stay small.

Usually, recovering a point parametrization and computing a
suitable knot vector are regarded as separate problems. Point parametrization
often is only analyzed in terms of interpolation which leads to an implicit knot
vector, whereas in the case of approximation, the knot vector usually is regarded as fixed. Other
methods focus on knot vector computation while choosing a well-known point
parametrization in a preprocessing step.

We propose a method that is able to simultaneously predict parametric values and knots using
two interdependent deep neural networks (DNN): (1)~a Point Parametrization Network (PPN) which assigns 
parametric values to point sequences; (2)~a Knot Selection Network (KSN) which predicts 
new knot values for knot vector refinement. Our neural networks directly operate on point 
data without the need for computing intermediate features. We will show that it is possible 
to include B-spline curve approximation directly into the network architecture.
When compared to state-of-the-art methods, our
parametrization leads to smaller approximation errors. To our knowledge, we are the first 
to show the potential of neural networks for
B-spline curve approximation.

The sections of this paper are arranged as follows. 
Section \ref{sec:relwork} presents related works.
Some required preliminaries are given in Section \ref{sec:prilim}.
Our parametrization approach is explained in Section \ref{sec:lppkp} while we
present our network architecture in Section \ref{DNNarc}.
Results and discussion can be found in Sections \ref{sec:results} and
\ref{sec:disc}.

\section{Related works} \label{sec:relwork}
Most knot placement methods require prior computation of parametric values.
Besides classic methods like the uniform parametrization and chord-length parametrization the most prominent one is the centripetal-method \cite{lee1989choosing}.
Another well-known approach is the universal method proposed by Lim \cite{lim1999universal} which leads to affine invariance and is closely related to the uniform method. The hybrid method by Shamsuddin et al. \cite{shamsuddin2004hybrid} is a mixture of the chord-length and centripetal method which leads to slightly higher accuracies.
Knot placement is usually an iterative process of inserting new knots until satisfaction of an error bound.
In \cite{li2005adaptive} a heuristic rule based on an angular measure is used to determine suitable knot values. An extensive analysis of the impact of different geometric features for knot vector computation is given by Razdan \cite{razdan1999knot}.
Piegl and Tiller \cite{piegl2012nurbs} average parametric values to generate the knot vectors. 
A well-known refinement based method was introduced by Park and Lee \cite{park2007b} where the knot vector relies on the computation of dominant points which are points of special interest (e.g. high curvature).
A machine learning approach using support vector machines for knot placement is described in \cite{laube2018learnt} which produces approximations with slightly higher error rates than the method by Park and Lee \cite{park2007b}.
Other methods use genetic algorithms for knot vector optimization \cite{valenzuela2013evolutionary, ulker2013b, tongur2016b} or meta-heuristics like a firefly algorithm driven approach \cite{galvez2013firefly}.

Machine learning has become an important field in geometry processing and has been successfully applied to different problems regarding geometric modeling.
Steinke et al.\ \cite{steinke2005support} use support vector regression for head reconstruction, outlier removal and hole filling.
Lin et al.\ \cite{lin2005neural} propose DNNs for surface reconstruction based on 2D images.
In \cite{qi2017pointnet} Qi et al.\ showed that DNNs can classify or segment point data without an intermediate representation. 
Another important research area is the application of machine learning in non-euclidean space, e.g., for shape-matching \cite{masci2015geodesic, boscaini2016learning} or shape-completion \cite{litany2017deformable}.

\section{Preliminaries} \label{sec:prilim}

In this section, we give a short introduction to B-spline curve approximation and deep neural networks.

\subsection{B-spline curve approximation}\label{sec:curvapr}

Suppose we have given a sequence of points $\*p=(p_0,...,p_m)$ with each point represented by coordinates $p_i = (x_i,y_i)$ as in Figures ... and ....
Consider a B-spline curve
$$
C(u) = \sum_{j=0}^n c_j \,N_{j}^{k}(u)
$$
of degree $k$ with control points $c_j$, B-spline functions $N_{i}^{k}(u)$, and a non-decreasing knot vector $\*u = (u_0,\dots,u_{n})$ where the knots $u_0$ and $u_n$ have multiplicity $k+1$ for end point interpolation.
To compute the control points $c_j$ of the B-spline curve $C$ approximating $\*p$, the least squares problem  
$$
\sum_{i=0}^m|p_i-C(t_i)|^2\to \min
$$
with precomputed parameters $t_i$, $i=0,\dots,m$ combined in the \emph{parameter vector} $\*t=(t_0,\dots,t_m)$ and end points $C(t_0)=c_0=p_0$ and $C(t_m)=c_n=p_m$ is solved.
This yields the normal equation
\begin{equation}
(\*N^T\*N)\*c = \*q \label{eq:normaleq}
\end{equation}
where $\*N$ is the $(m-1) \times (n-1)$ matrix
$$
\*N = \begin{pmatrix}
N_{1}^{k}(t_1) & \dots & N_{n-1}^{k}(t_1) \\
\vdots & \ddots & \vdots \\
N_{1}^{k}(t_{m-1}) & \dots & N_{n-1}^{k}(t_{m-1}) 
\end{pmatrix},
$$
and $\*c$ and $\*q$ are the vectors defined as
$$
\*c = \begin{pmatrix}
c_1 \\
\vdots \\
c_{n-1}
\end{pmatrix},
\*q = \begin{pmatrix}
\sum_{i=1}^{m-1} N_{1}^{k}(t_i)q_i \\
\vdots \\
\sum_{i=1}^{m-1} N_{n-1}^k(t_i)q_i \\
\end{pmatrix}
$$
and 
\begin{equation*}
q_i = p_i-N_{0}^{k}(t_i)p_0-N_{n}^{k}(t_i)p_m 
\end{equation*}
for $i=1,...,m-1$.
If there are no constraints for end point interpolation (\ref{eq:normaleq}) reduces to 
 \begin{equation}
(\*N^T\*N)\*c = \*N^T\*p \label{eq:normaleqnoconstr}.
\end{equation}
The control points $c_j$ can be computed using (\ref{eq:normaleq}), if
\begin{equation}
	\sum_{l=1}^{m-1} N^k_i(t_l)N^k_j(t_l)\ne 0. \label{eq:SW}
\end{equation}
This is equivalent to the existence of a parameter $t_i \in [u_j,u_{j+1}]$ for $j=k,...,n+1$, see e.g.\ \cite{farin2002curves}.
For our experiments, we used $k=3$.

\subsection{Deep neural networks }\label{sec:dnn}
In this work, we apply feed-forward DNNs to learning the parameter vector
$\*t$ for points $\*p$ and a knot vector $\*u$ to get tight B-spline curve approximations. 
DNNs are organized in layers of uniformly behaving artificial neurons. Some of these layers are 
trainable, i.e., their behavior is controlled by adjustable weights. 
Here, we use the classical multilayer perceptron (MLP) architecture (Figure \ref{fig:paramnet}) where each neuron 
is connected via weights to all neurons of the previous layer. 
Each neuron computes a weighted sum of its inputs and feeds it through a static nonlinearity as its output which
again is connected to all neurons in the next layer. 
The weights are adapted during the training phase by gradient descent on a loss function that measures the 
performance of the DNN (Section \ref{sec:traindat}). 
The reader is referred to Goodfellow et al.\ \cite{goodfellow2016deep} for an in-depth discussion of deep learning.

In order to apply DNNs to our problem of parametrization, we have to address the following challenges:

\begin{enumerate}[label=CL\arabic*.]
	\item Since there are no publicly available datasets for this problem we have to 
	synthesize a sufficiently large training data set.
	\item It must be ensured that the training data and the real data share the same 
	characteristics.
	\item A suitable loss function for a parametrization of a B-spline
	approximation must be defined.
	\item Due to its architecture, a MLP requires input of a fixed size whereas point sequences for approximation are of variable size. The approach must be able to cope with this problem.
\end{enumerate}

\section{Learning Point Parametrization and Knot Placement}\label{sec:lppkp}
\begin{figure*}[htb!]
\centering
\includegraphics[width=0.9\textwidth]{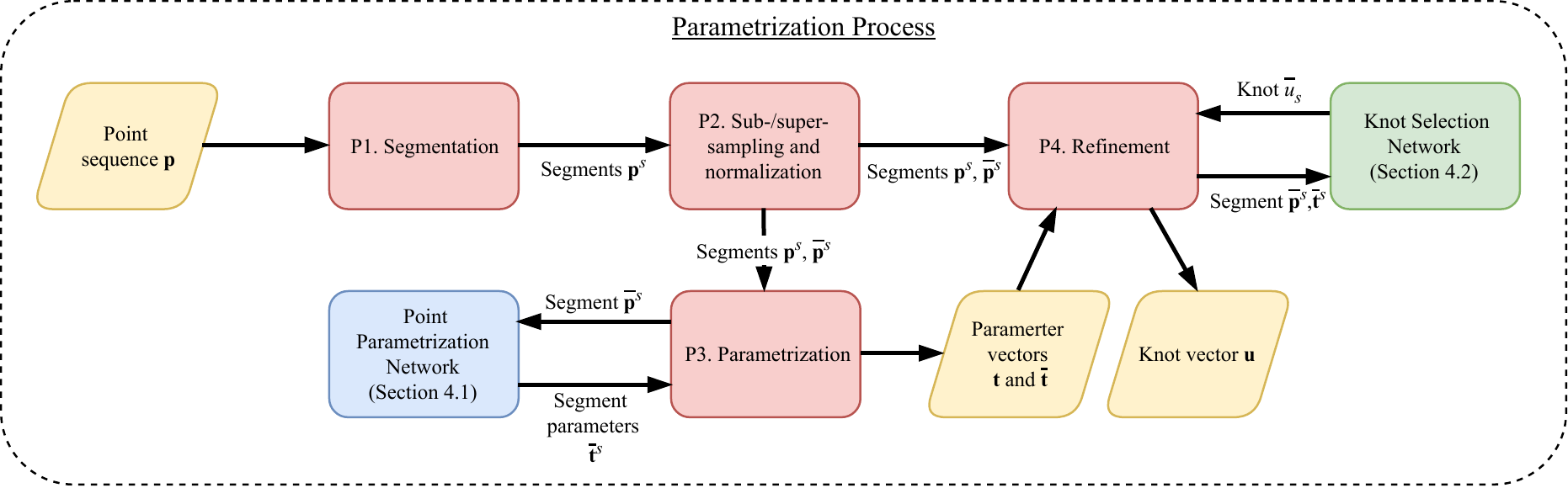}
\caption{Overview of the parametrization process.
The inputs/outputs are marked in yellow.
Red boxes refer to sub-processes described successively in Section \ref{sec:lppkp}. 
The other colors refer to the sub-processes in Figure \ref{fig:paramnet}.}
\label{fig:paramnetwf}
\end{figure*}

In this section we will describe our method for point parametrization and knot placement for arbitrary 2D point sequences $\*p$.
In our exposition we follow Figure \ref{fig:paramnetwf} which gives an overview of the parametrization process.\\
\\
\textbf{P1.\ Segmentation}  
The input point sequence $\*p$ has a complexity given by its total curvature
$$
	\hat{\kappa}(\*p) = \sum_{i=0}^{m-1} \frac{(|\kappa_i| + |\kappa_{i+1}|)\|p_{i+1}-p_i\|_2}{2},
$$
where $\kappa_i$ is the curvature at point $p_i$.
Throughout this work curvature is computed using osculating circles \cite{taubin1991estimation}.
Using $\hat{\kappa}(\*p)$ as a measure of complexity we can quantify the maximum complexity of the training data set and by that the maximum complexity the DNN is able to process.
To handle CL2. we propose to split point sequences so that the resulting segments represent the complexity of the training data. 
For a point sequence $\*p$ compute the total curvature $\hat{c}(\*p)$ and split $\*p$ into point sequence segments $\*p^s, s=1,...,r,$ at the median, if $\hat{\kappa}(\*p)>\hat{\kappa}_t$ for a threshold $\hat{\kappa}_t$. We set $\hat{\kappa}_t$ to the $98$th percentile of $\hat{\kappa}(\cdot)$ of the training set (Section \ref{sec:traindat}). 
We repeat this segmentation process $r-1$ times until each segment $\*p^s$ satisfies $\hat{\kappa}(\*p^s) < \hat{\kappa}_t$.

\paragraph{P2.\ Sub-/supersampling and normalization} 
To process $\*p^s, s=1,...,r,$ by the PPN/KSN the number of points per segment has to match the DNN input size $l$ (CL4.).
Thus, the $\*p^s$ are sub- or supersampled:
\begin{description}
	\item[Subsampling:] If the number of points in $\*p^s$ is larger than $l$, draw points from $\*p^s$ such that the drawn indices $i$ are equally distributed and include the first and last point. 

	\item[Supersampling:] If the number of points in $\*p^s$ is smaller than $l$, we linearly interpolate  temporary points between consecutive points $p_i^s$ and $p_{i+1}^s$ from left to right.
This is iterated until the number of points equals $l$.
These temporary points are not permanent members of $\*p$.
They only exist to matching the network input size.
\end{description}
The sub-/supersampled segments are then normalized to $\bar{\*p}^s$ consisting of the points
$$
	\bar{p}_i^s = \frac{p_i^s- \min(\*p^s)}{\max(\*p^s)-\min(\*p^s)},
$$
where $\min(\*p^s)$ and $\max(\*p^s)$ are the minimum and maximum coordinates of $\*p^s$.

\paragraph{P3.\ Parametrization} 
For each $\bar{\*p}^s$ the PPN (see Section \ref{PPN}) generates a parametrization $\bar{\*t}^s\subset[0,1]$. 
This parametrization is rescaled to $[u_{s-1},u_s]$ and adapted to the sub-/supersampling of $\bar{\*p}^s$, yielding $\*t^s$.
For a point $p_i$ that was removed from $\*p^s$ in the subsampling,
insert to $\bar{\*t}^s$ a parameter 
$$
	t_i = t_{\alpha}^s + (t_{\omega}^s - t_{\alpha}^s)\frac{\arclen(p_{\alpha}^s, p_i)}{\arclen(p_{\alpha}^s, p_{\omega}^s)},  
$$
where $\arclen$ is the length of the polygon defined by a point sequence and
$p_{\alpha}^s$ and $p_{\beta}^s$ are the closest neighbors of $p_i$ to the left and right in the subsampled segment with parameters $t_{\alpha}^s$ and $t_{\beta}^s$.
Parameters $t_i^s$ corresponding to temporary points are simply removed from $\bar{\*t}^s$.

For the initialization of the parametrization step, an initial knot vector is required.
First define $u_0=0$ and $u_n=1$.
Then, for each segment (except the last one), one knot $u_i$ is added
$$
	u_i = u_{i-1} + \frac{\arclen(\*p^s)}{\arclen(\*p)}, \quad i=1,...,r-1.
$$
This yields a start- and end-knot for every point sequence segment.

\paragraph{P4.\ Refinement}  
In the refinement step, additional knots are added to point sequence segments with large approximation error.
We determine the segment $\*p^s$ with largest Hausdorff distance to the input data $\*p$.
For $\bar{\*p}^s$ and $\bar{\*t}^s$ the KSN (see Section \ref{KSN}) generates a new estimated knot $\bar{u}_s\in[0,1]$.
The new knot $\bar{u}_s$ is mapped to the actual knot value range $[u_{s-1}, u_{s}]$ by 
$$
	\widetilde{u}_s=u_{s-1} + \bar{u}_s(u_{s}-u_{s-1}).
$$
Then, instead of $\widetilde{u}_s$, the parameter value $t_i$ closest to $\widetilde{u}_s$ is inserted to $\*u$.
Since the KSN operates on sub-/supersampled data, this correction of $\widetilde{u}_s$ is the simplest choice to ensure (\ref{eq:SW}). 
\begin{remark}\label{rem}
Note, that $\widetilde{u}_s$ could be inserted to $\*u$ directly, as long as (\ref{eq:SW}) is satisfied.
\end{remark}
We propose to further refine $\*u$ until the desired curve approximation error threshold is satisfied.

\section{DNN architectures} \label{DNNarc}
\begin{figure*}[htb!]
\centering
\includegraphics[width=0.9\textwidth]{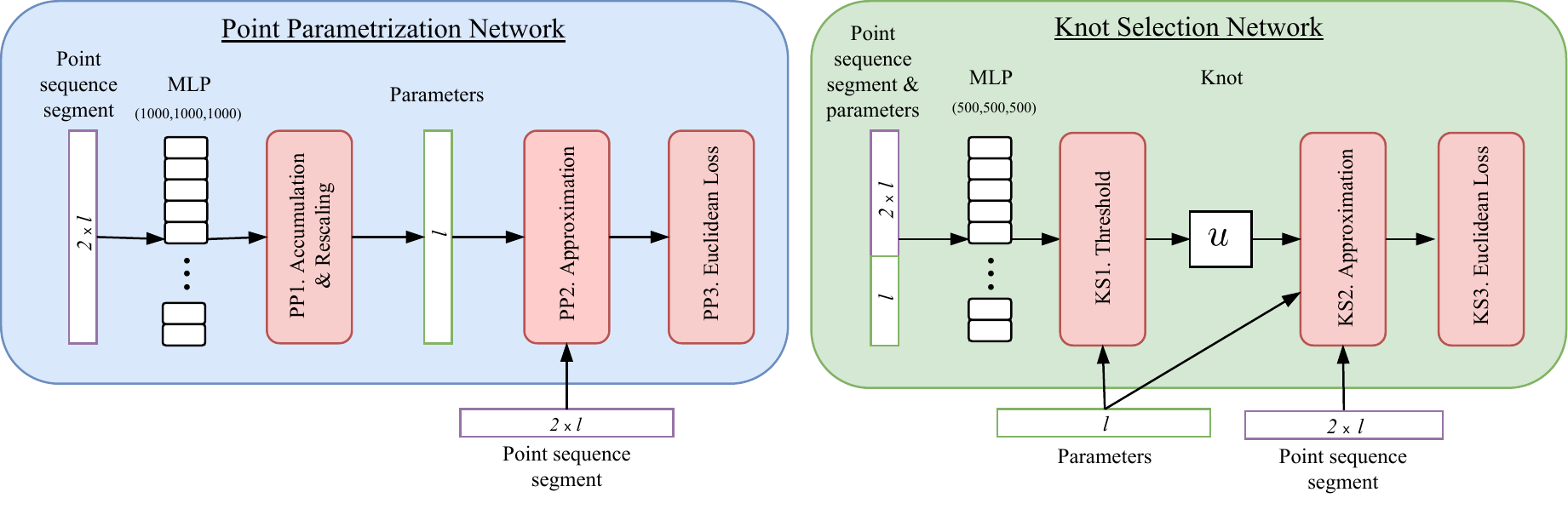}
\caption{Network architectures for the Point Parametrization Network (PPN, left) and the Knot Selection Network (KSN, right).
Red boxes refer to sub-processes described successively in Sections \ref{PPN} and \ref{KSN}.}
\label{fig:paramnet}
\end{figure*}

In this section we describe the deep neural networks architectures for point 
parametrization (PPN) and knot selection (KSN) and their training.
Figure \ref{fig:paramnet} shows the network architectures of these networks.
Since these networks take point sequence segments $\bar{\*p}^s$ as input we will drop 
the upper index $s$ and the over-bar for all variables in Sections \ref{PPN} and \ref{KSN} 
to simplify notation.

\subsection{Point Parametrization Network} \label{PPN}
For a sequence of points $\*p$ a parameter vector $\*t=(t_i)_i$ is defined as $t_i = t_{i-1} + \Delta_{i-1}$. 
For classical methods $\Delta_i$ is computed based on geometric properties of $\*p$, e.g.\ the  centripetal parametrization $\| p_{i+1} - p_i\|^{\frac{1}{2}}$.
We propose to estimate $\Delta_i$ using a pre-trained neural network called the Point Parametrization Network. 
Similarly, the input to the PPN consists of segments $\*p$.
It can be written in the form $\*p = (x_0,...,x_{l-1},y_0,...,y_{l-1})$, where the $x$ and $y$ are the coordinates of the points of $\*p$. 
The parameter domain is defined as $u_0 = t_0 = 0$ and $u_n = t_{l-1} = 1$. 
Then, the task of the PPN is to predict missing values $\Delta = (\Delta_0, ..., \Delta_{l-2})$ with 
\begin{equation}
	\Delta_i > 0, i=0,\dots,l-2,
	\label{eq:deltat}
\end{equation}
such that $t_0 < t_1$ and $t_{l-2} < t_{l-1}$. 
We apply a MLP to the input data $\*p$, yielding as output a 
distribution for parametrization 
$\Delta^{mlp}=(\Delta^{mlp}_0, ..., \Delta^{mlp}_{l-2})$ of size $l-1$.

\paragraph{PP1.\ Accumulation \& Rescaling}  
The output $\Delta^{mlp}$ is used to compute a parameter vector $\*t^{mlp}$ with $t_0^{mlp}=0$ and 
$$
	t_i^{mlp} = \sum_{j=0}^{i-1} \Delta^{mlp}_j,
	i=1,\dots,l-1. 
$$
Since, $t_{l-1}^{mlp}$ is usually not $1$, rescaling of $\*t^{mlp}$ yields the final parameter vector $\*t$ with
$$
	t_i = {t^{mlp}_i}{/\max(\*t^{mlp})}.
$$
To ensure (\ref{eq:deltat}), the MLP output $\Delta^{mlp}$ must be positive.
This is achieved by using the softplus activation function
$$
f(x) = \ln(1+e^x)
$$
for neurons of the MLP in the PPN.

\paragraph{PP2.\ Approximation}
To be able to define a network loss (CL3.) we include the B-spline curve approximation directly as a network layer. 
The input points $\*p$ and their parameters $\*t$ are used for an approximation with knot vector 
$\*u = (0,0,0,0,1,1,1,1)$ for $k=3$. 
Since the PPN parametrizes curve segments and not the complete curve we approximate without endpoint interpolation (\ref{eq:normaleqnoconstr}). 
The approximation layer's output $\*p^{app}=(p^{app}_0,\dots,p^{app}_{l-1})$ is the approximating B-spline curve evaluated at $\*t$. 

\paragraph{PP3.\ Euclidean Loss} The loss for the PPN is 
\begin{equation}
	\frac{1}{l} \sum_{i=0}^{l-1} \|p_i - {p}_i^{app}\|_2.
	\label{eq:loss}
\end{equation}

\subsection{Knot Selection Network} \label{KSN}
The KSN predicts a new knot $u$ to the interval $(0, 1)$ for 
a given segment $\*p$ and parameters $\*t$ (predicted by the PPN).
Thus, the resulting network input size is $3l$.

We apply a MLP, which transforms the input to a single output value ${u}^{mlp}$.
As network activation functions we use the RELU function \cite{glorot2011deep} except for the output layer, where we use the Sigmoid function \cite{han1995influence}.

\paragraph{KS1.\ Threshold Layer}
The new knot $u$ has to satisfy 
$u \in(0, 1)$, and
$\*t\cap[0,u]\ne\emptyset$ and 
			$\*t\cap[u,1]\ne\emptyset$, to ensure (\ref{eq:SW}).
To satisfy these constraints, we use a threshold layer which maps ${u}^{mlp}$ to 
$$
{u} = 
\begin{cases}
    \epsilon & \text{, if } {u}^{mlp} \leq 0 \\
    1 - \epsilon & \text{, if } {u}^{mlp} \geq 1 \\
    u^{mlp}            & \text{, otherwise}.
\end{cases}
$$
By introducing a small $\epsilon = 1\mathrm{e}{-5}$ we make sure that knot multiplicity at the end-knots stays equal to $k$.
This choice of $u$ corresponds to the more general approach mentioned in Remark \ref{rem}. 

\paragraph{KS2.\ Approximation} The approximation in the KSN has the same form as the approximation of the PPN with the exception of the knot vector. 
Here, the knot vector is $\*u = (0,0,0,0,{u},1,1,1,1)$. 
For backpropagation, the derivative of the B-spline basis functions with respect to ${u}$ is required, see \cite{piegl1998computing}. 
As for the PPN, the approximation layer's output $\*p^{app}=(p^{app}_0,\dots,p^{app}_{l-1})$ is the approximating B-spline curve evaluated at $\*t$. 

\paragraph{KS3.\ Euclidean Loss} The loss function for the KSN is the same as for the PPN, see (\ref{eq:loss}).

\subsection{Training set generation and training process} \label{sec:traindat}

For the training of the PPN and the KSN sufficiently large training and test datasets are required.
As for the input size we define $l=100$.
An ideal dataset would consist of diverse real-world point clouds, with a known sequential 
order of points.
Since no such datasets are publicly available, we chose to synthesize the data using 
B-spline curves.
We generate random control points $c_i$ using a normal distribution with mean $\mu$ and 
variance $\sigma$ to define B-spline curves of degree $k=3$ with $(k+1)$-fold end-knots 
and no interior knots.
For the $y$-coordinates, we use $\sigma = 2$ and $\mu = 10$.
For the $x$-coordinates, we use $\sigma = 1$ and $\mu = 10$
for the first control point and increase $\mu$ by $\Delta\mu=1$ for all consecutive 
control points. 
Curves with self-intersections are discarded, because the sequential order of their 
sampled points is not unique and, in reverse engineering, such point sets are usually 
split into subsets at the self-intersection.
Smaller $\sigma$ for control point $x$-coordinates reduces the number of curves with self-intersections in the first place. 
Using this approach, we generate a dataset consisting of $150.000$ curves. Then, we sample $l$ points $\*p=(p_0,\dots,p_{l-1})$ along each curve.
Since these curves tend to have increasing x-coordinates from left to right we add 
index-flipped versions of the point sequences to the dataset resulting in $300.000$ point 
sequences of which 20\% are used as test data in the training process.
In our experiments, this method leads to a very diverse set of curves containing sections 
with very little up to no curvature as well as sections with high curvature and even sharp 
features.
While we use cubic B-spline curves for dataset generation, our approach is not limited to 
point clouds computed this way.

\begin{figure}[htb]
\centering
\includegraphics[width=0.7\columnwidth]{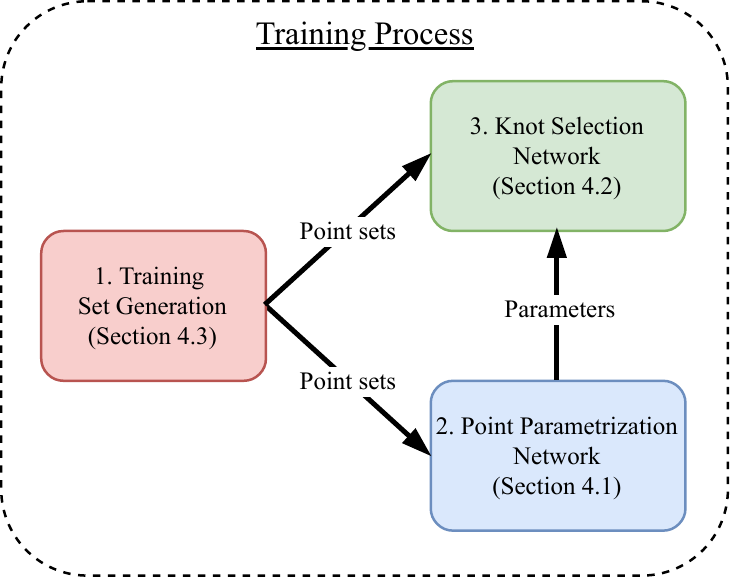}
\caption{Overview of the training process.
The red box refers to the training set generation in Sec. \ref{sec:traindat}. 
The other colors refer to the neural networks described in Sec. \ref{sec:lppkp}.}
\label{fig:trainnetwf}
\end{figure}

Since the KSN requires point parametrizations $\*t$ as input, the PPN is trained first.
After training, the layers PP2 and PP3 are discarded and PP1 becomes the network output layer.
We compute parametric values $\*t$ for the training dataset by applying the PPN and train 
the KSN on the combined input. For parametrization in Section \ref{sec:lppkp}, the layers 
KS2 and KS3 are discarded while KS1 becomes the network output layer.
The MLPs of the PPN and KSN consist of three hidden layers with sizes $(1000,1000,1000)$ 
and $(500,500,500)$.
Figure \ref{fig:trainnetwf} gives an outline of the training process.
We apply dropout \cite{hinton2012improving} to MLP layers and train the networks using the
Adam optimizer \cite{kingma2014adam}.

\section{Results}\label{sec:results}

In this section we present results of our parametrization method which we call
PARNET.
First, we discuss results of the point parametrizations computed by the PPN
(Section \ref{eval:PP}).
Then, we discuss knot selections computed by the KSN as well as the global
approximation quality of our approach (Section \ref{eval:KS}).

For the evaluation we generated four evaluation sets:
\begin{itemize}
    \item \textit{Evaluation set 1} contains $500$ curves computed as described in Section
    \ref{sec:traindat}. We sample $500$ equidistributed (in terms of arc length) points on each curve.
    \item \textit{Evaluation set 2} contains the curves from \textit{evaluation set 1} but
    sampled at random parameters.
    \item \textit{Evaluation set 3} contains $500$ curves computed as described in Section
    \ref{sec:traindat} but with random interior knots without multiplicities. 
		We generate 
    $3$ to $8$ random interior knots which results in a set of very diverse curves, some of 
    high complexity. We sample $500$ equidistributed points on each
    curve.
    \item \textit{Evaluation set 4} contains the curves from \textit{evaluation set 3} but
    sampled at random parameters.
\end{itemize}
We included \textit{evaluation set 2} and \textit{evaluation set 4} into our
evaluation because many parametrization methods use noise filters before parametrization, 
e.g.\ \cite{schall2005robust}.  
These filters result in a smooth set of points (or smooth 
curvature) but also lead to an uneven distribution of points. 
Training, test and 
evaluation sets can be downloaded from (http://www.ios.htwg-konstanz.de/parnetdatasets).

\subsection{Point Parametrization} \label{eval:PP}
\begin{table}
\centering
\begin{tabular}{l*{2}|c|c}
& \textit{Evaluation set 1} & \textit{Evaluation set 2}\\
\hline
PNN 		      	& 0.0224 & 0.0992  \\
Uniform         & 0.2097 & 0.2095  \\
Chordal         & 0.2099 & 0.2001  \\
Centripetal     & 0.2098 & 0.2030  \\
\hline
PPN             & 0.0245 & 0.1088 \\
uniform         &  0.2040 & 0.2102 \\
chordal         &  0.2042 & 0.1955 \\
centripetal     & 0.2040  & 0.2008 \\
\end{tabular}
\caption{Average Hausdorff distances for approximation without interior knots of
the PPN compared to common parametrization methods.}
\label{tab:1}
\end{table}
\begin{figure}[htb!]
    \centering
    \subfloat[]{
    \includegraphics[width=0.15\textwidth]{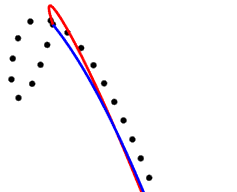}
    \label{fig:1p}
    }
    \subfloat[]{
    \includegraphics[width=0.13\textwidth]{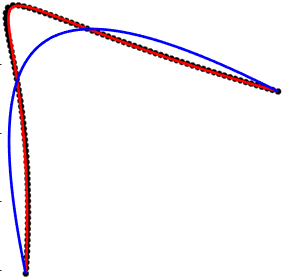}
    \label{fig:2p}  
    }
    \subfloat[]{
    \includegraphics[width=0.15\textwidth]{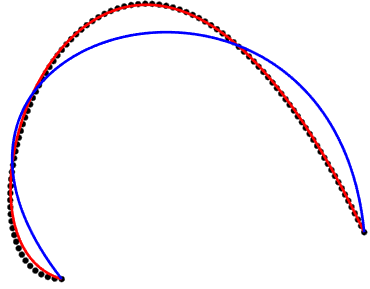}
    \label{fig:2p}  
    }    
    \newline
    \subfloat[]{
    \includegraphics[width=0.3\textwidth, height=2cm]{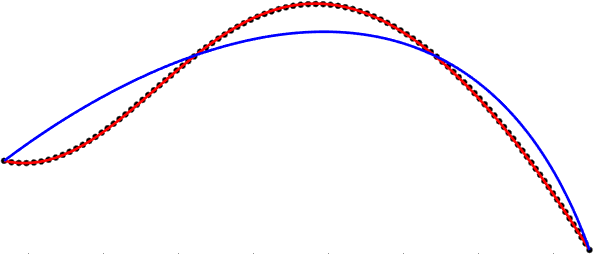}
    \label{fig:3p}  
    }
    \caption{Results of parametrizations for examples of \textit{evaluation set 1} by the centripetal method (blue) and by the PPN (red), approximated without interior knots.}
\label{fig:paramsamp}
\vspace{-1em}
\end{figure}

\begin{figure}[htb!]
\centering
\includegraphics[width=0.38\textwidth]{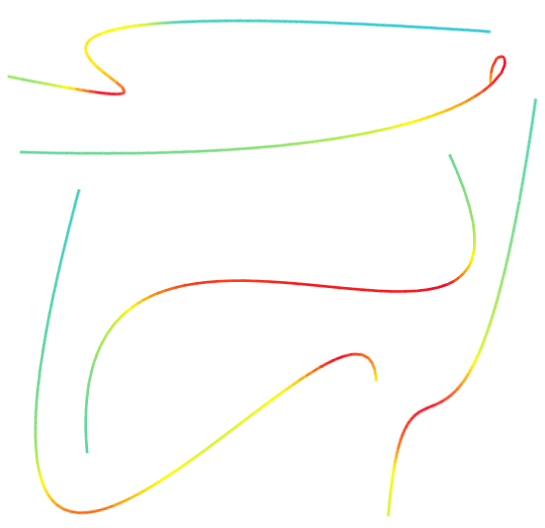}
\caption{Heatmap-colored curves of \textit{evaluation set 3} colored by
parametrization value $\Delta$. 
Blue corresponds to low values for $\Delta$ while red corresponds to large
values for $\Delta$.}
\label{fig:testset}
\end{figure}

In Table \ref{tab:1} we compare point parametrizations computed by the PPN and
the uniform, chordal, and centripetal parametrization. For evaluation of curve
approximation quality, Hausdorff distance is the de facto standard
\cite{sklansky1980fast, chen2010cubic}.
We compare the methods for equi-distributed as well as randomly sampled points
in terms of the average Hausdorff distance over the complete evaluation set.
Parametrizations computed by the PPN result in approximations with up to eight
times smaller Hausdorff distance for \textit{evaluation set 1}. 
For \textit{evaluation set 2} the results are still two times smaller when
compared to the other methods.  
Most methods for parametrization are based on geometric relations of points. 
It has been shown that high curvature is a strong indicator for a denser
parametrization \cite{ma1995parameterization}.
In \cite{lee1989choosing}, Lee introduces the general exponent method which also
includes the centripetal parametrization. 
Here parameter values are based on changes in curvature. 
Assuming that regions of higher curvature are sampled more densely, this method
fails for equidistributed points as can be seen from results in Table
\ref{tab:1} and Figure \ref{fig:paramsamp}.
As can be seen in Figure \ref{fig:paramsamp} approximations using our method are able to follow the points very closely. Figure \ref{fig:1p} shows a close-up on an example where the PPN results in larger Hausdorff distance when compared to the centripetal method.
In Figure \ref{fig:testset} we colored curves from \textit{evaluation set 1}
according the distribution of the parametrization $\Delta$.
It is obvious, that a large absolute curvature is a strong indicator for
larger values in $\Delta$. 
But also regions containing inflection points lead to large values in $\Delta$. 
Since points in \textit{evaluation set 3} are equidistributed, the PPN has
learned to incorporate curvature into the parametrization process.

\begin{figure}[htb!]
\centering
\scalebox{0.7}{
%
%
\begin{tikzpicture}

\begin{axis}[%
width=2.5in,
height=2in,
at={(0.758in,0.481in)},
scale only axis,
xmin=3,
xmax=23,
xlabel={number of knots},
ymin=0,
ymax=0.4,
ytick={   0, 0.05, 0.1, 0.15, 0.2, 0.25, 0.3, 0.35, 0.4},
scaled y ticks = false,
y tick label style={/pgf/number format/fixed,
      /pgf/number format/1000 sep = \thinspace 
      },
ylabel={average Hausdorff distance},
axis background/.style={fill=white},
axis x line*=bottom,
axis y line*=left,
legend style={at={(0.97,0.78)},anchor=east,legend cell align=left,align=left,draw=white!15!black}
]
\addplot [color=orange,solid,line width=1.5pt]
  table[row sep=crcr]{%
3	0.3698\\
5	0.2481\\
7	0.1787\\
9	0.1421\\
11	0.1151\\
13	0.0955\\
15	0.0838\\
17	0.0704\\
19	0.0637\\
21	0.0554\\
23	0.0512\\
};
\addlegendentry{NKTP};

\addplot [color=blue,solid,line width=1.5pt]
  table[row sep=crcr]{%
3	0.4488\\
5	0.2374\\
7	0.1314\\
9	0.0919\\
11	0.0507\\
13	0.0363\\
15	0.0275\\
17	0.019\\
19	0.0191\\
21	0.016\\
23	0.014\\
};
\addlegendentry{DPKP};

\addplot [color=green,solid,line width=1.5pt]
  table[row sep=crcr]{%
3	0.3272\\
5	0.1689\\
7	0.1063\\
9	0.0781\\
11	0.0508\\
13	0.039\\
15	0.03\\
17	0.024\\
19	0.02\\
21	0.016\\
23	0.013\\
};
\addlegendentry{PARNET};

\addplot [color=black,solid,forget plot]
  table[row sep=crcr]{%
3	0.4\\
23	0.4\\
};
\addplot [color=black,solid,forget plot]
  table[row sep=crcr]{%
3	0.35\\
23	0.35\\
};
\addplot [color=black,solid,forget plot]
  table[row sep=crcr]{%
3	0.3\\
23	0.3\\
};
\addplot [color=black,solid,forget plot]
  table[row sep=crcr]{%
3	0.25\\
23	0.25\\
};
\addplot [color=black,solid,forget plot]
  table[row sep=crcr]{%
3	0.2\\
23	0.2\\
};
\addplot [color=black,solid,forget plot]
  table[row sep=crcr]{%
3	0.15\\
23	0.15\\
};
\addplot [color=black,solid,forget plot]
  table[row sep=crcr]{%
3	0.1\\
23	0.1\\
};
\addplot [color=black,solid,forget plot]
  table[row sep=crcr]{%
3	0.05\\
23	0.05\\
};
\end{axis}
\end{tikzpicture}%
}
\caption{Average Hausdorff distance over \textit{evaluation set 3} at different
numbers of knots for PARNET, DPKP and NKTP}
\label{fig:hd}
\end{figure}
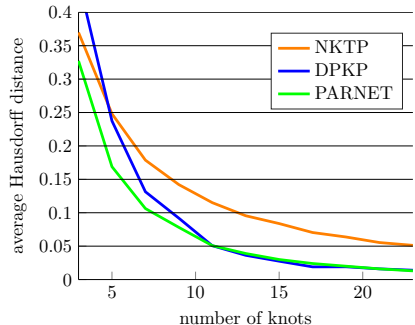
\begin{figure}[htb!]
\centering
\scalebox{0.7}{
%
%
\begin{tikzpicture}

\begin{axis}[%
width=2.5in,
height=2in,
at={(0.758in,0.481in)},
scale only axis,
xmin=3,
xmax=23,
xlabel={number of knots},
ymin=0,
ymax=0.4,
ytick={   0, 0.05, 0.1, 0.15, 0.2, 0.25, 0.3, 0.35, 0.4},
scaled y ticks = false,
y tick label style={/pgf/number format/fixed,
      /pgf/number format/1000 sep = \thinspace 
      },
ylabel={average Hausdorff distance},
axis background/.style={fill=white},
axis x line*=bottom,
axis y line*=left,
legend style={at={(0.97,0.78)},anchor=east,legend cell align=left,align=left,draw=white!15!black}
]
\addplot [color=orange,solid,line width=1.5pt]
  table[row sep=crcr]{%
3	0.3759\\
5	0.2462\\
7	0.1811\\
9	0.1436\\
11	0.1178\\
13	0.099\\
15	0.085\\
17	0.0760\\
19	0.068\\
21	0.059\\
23	0.054\\
};
\addlegendentry{NKTP};

\addplot [color=blue,solid,line width=1.5pt]
  table[row sep=crcr]{%
3	0.6228\\
5	0.4761\\
7	0.3517\\
9	0.2307\\
11	0.1473\\
13	0.091\\
15	0.0605\\
17	0.0438\\
19	0.033\\
21	0.027\\
23	0.023\\
};
\addlegendentry{DPKP};

\addplot [color=green,solid,line width=1.5pt]
  table[row sep=crcr]{%
3	0.3362\\
5	0.2016\\
7	0.1259\\
9	0.0922\\
11	0.071\\
13	0.06\\
15	0.0472\\
17	0.042\\
19	0.0347\\
21	0.03\\
23	0.027\\
};
\addlegendentry{PARNET};

\addplot [color=black,solid,forget plot]
  table[row sep=crcr]{%
3	0.4\\
23	0.4\\
};
\addplot [color=black,solid,forget plot]
  table[row sep=crcr]{%
3	0.35\\
23	0.35\\
};
\addplot [color=black,solid,forget plot]
  table[row sep=crcr]{%
3	0.3\\
23	0.3\\
};
\addplot [color=black,solid,forget plot]
  table[row sep=crcr]{%
3	0.25\\
23	0.25\\
};
\addplot [color=black,solid,forget plot]
  table[row sep=crcr]{%
3	0.2\\
23	0.2\\
};
\addplot [color=black,solid,forget plot]
  table[row sep=crcr]{%
3	0.15\\
23	0.15\\
};
\addplot [color=black,solid,forget plot]
  table[row sep=crcr]{%
3	0.1\\
23	0.1\\
};
\addplot [color=black,solid,forget plot]
  table[row sep=crcr]{%
3	0.05\\
23	0.05\\
};
\end{axis}
\end{tikzpicture}%
}
\caption{Average Hausdorff distance over \textit{evaluation set 4} at different
numbers of knots for PARNET, DPKP and NKTP}
\label{fig:testset_uneven}
\end{figure}
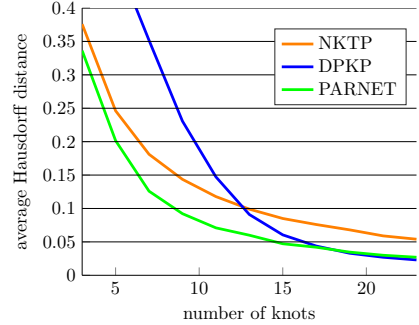
\subsection{Knot Selection} \label{eval:KS}
\begin{figure}[htb!]
    \centering
    \subfloat[]{
    \includegraphics[width=0.06\textwidth]{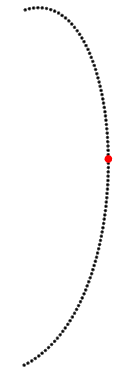}
    \label{fig:1k}
    }
    \subfloat[]{
    \includegraphics[width=0.13\textwidth]{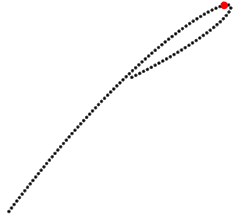}
    \label{fig:2k}  
    }
    \subfloat[]{
    \includegraphics[width=0.12\textwidth]{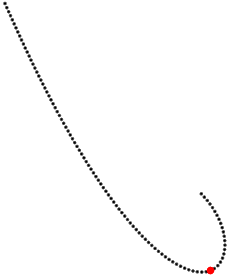}
    \label{fig:3k}  
    }
    \subfloat[]{
    \includegraphics[width=0.13\textwidth]{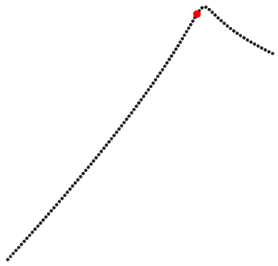}
    \label{fig:4k}
    }\newline
    \subfloat[]{
    \includegraphics[width=0.1\textwidth]{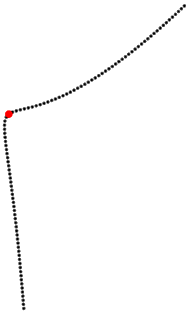}
    \label{fig:5k}  
    }
    \subfloat[]{
    \includegraphics[width=0.11\textwidth]{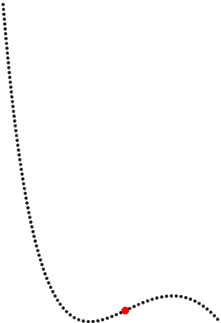}
    \label{fig:6k}  
    }
    \subfloat[]{
    \includegraphics[width=0.145\textwidth]{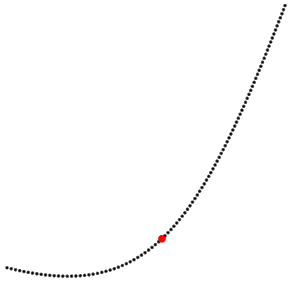}
    \label{fig:7k}
    }
    \subfloat[]{
    \includegraphics[width=0.095\textwidth]{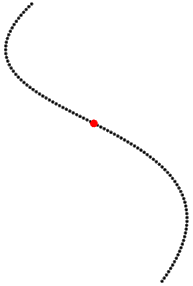}
    \label{fig:8k}  
    }
    \caption{Single knot (red) selected by the KSN on samples of
\textit{evaluation set 1}.}
\label{fig:singeknot}
\vspace{-1em}
\end{figure}

We evaluate the effectiveness of our approach by comparing PARNET to two other methods
for knot placement. One is a the well-known averaging method by Piegl and Tiller
\cite{piegl1998computing} which does not incorporate any geometric information
in the process of knot placement (NKTP). Since we place knots using refinement
we also compare PARNET to a well-known refinement-based method by Park et al.
\cite{park2007b} which uses so-called dominant points for knot placement (DPKP).
Again, methods are compared by average Haussdorf distance. Figures \ref{fig:testset} and
\ref{fig:testset_uneven} show the results of knot placement in a range from $3$
to $23$ knots on the \textit{evaluation sets} \textit{1} and \textit{2}.
On both sets, our method produces approximations of higher quality. Especially
with fewer knots in the range from $3$ to $12$ knots, our method has
significantly lower Haussdorf distance. 
With an increasing number of knots, results of DPKP
and our method are very close with a small advantage for our method at $23$
knots on \textit{evaluation set 3} and for DPKP on \textit{evaluation set 4}.
Figure \ref{fig:curvelarge} shows an approximation by the different methods for one
example from \textit{evaluation set 3}.
 While NKTP produces very smooth results
it fails in complex regions. 
The DPKP method is able to approximate regions of
high curvature very well but may also lead to wiggles in these regions. Our
method is able to approximate highly curved regions while also producing smooth
approximations (see the supplementary material for more examples).
In Figure \ref{fig:singeknot}, we present some curves of \textit{evaluation set
1} and predicted knot positions for refinement by the KSN. 
Examining knot predictions made by the KSN we make several observations that
agree with observations of other authors:
\begin{figure*}[htb!]
\centering
\includegraphics[width=0.8\textwidth]{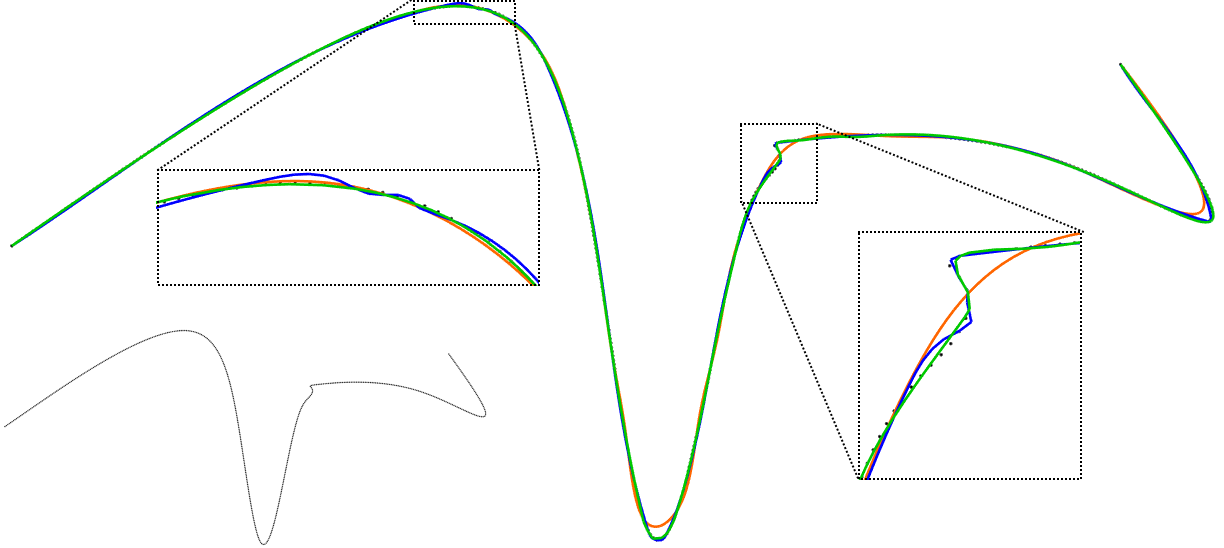}
\caption{Approximation results with $23$ knots for one example of
\textit{evaluation set 3} by methods NKTP (orange), DPK (blue) and PARNET
(green). 
The original point sequence is shown in the bottom left corner while
the framed boxes contain close-ups of critical regions.}
\label{fig:curvelarge}
\end{figure*}
\begin{itemize}
\item Curvature plays an important role in knot placement \cite{razdan1999knot,
yuan2013adaptive}. Regions of high curvature should be favored when placing
knots (see Figures \ref{fig:2k}, \ref{fig:3k}, \ref{fig:4k}, and \ref{fig:5k}).
\item If one has to refine segments with varying curvature directions placing the knot
at an inflection point is beneficial \cite{laube2018learnt, park2007b}. 
Examples
of the KSN choosing to place the knot near an inflection point and not at high
curvature regions can be seen in Figures \ref{fig:6k} and \ref{fig:8k}.
\item If curvature is small or changes slowly it is beneficial to split segments so that
resulting segments are of equal complexity \cite{park2007b, razdan1999knot}.
This can be seen in Figures \ref{fig:1k} and \ref{fig:7k} where the KSN splits the
point sets close to the median index but makes segments containing higher
curvature smaller. 
\end{itemize}

\section{Discussion} \label{sec:disc}
Our experiments show that neural networks are able to successfully predict simultaniously parametric values $\*t$ and knots $\*u$ for the problem of B-spline curve approximation. Our method results in tight approximations. It works well for unevenly spaced points although we trained on evenly spaced point sequences. This shows that our network is able to generalize well to previously unseen data with versatile characteristics. One aspect that limits our approach is the synthetic training data set. 
Real world data would be preferable. Since it is common to retrain networks for the purpose of specialisation our approach can be used as a pre-training method, which may be subsequently improved on additional data. Another drawback of our method is the need for segmentation as well as sub- and supersampling of point sequences. We deliberately chose to segment and sample in a very simple fashion to show that the approximation quality is not a result of preprocessing but of the parametrization by the networks.
In our approach B-spline curve approximation is directly integrated into the network training loop.
We hope that this will enable others to apply neural networks for approximation-related problems.
For future work we plan to investigate methods such as recurrent networks with attention in order to become more independent of segmentation and sampling. We also would like to apply our approach to surface approximation and other parametric representations like T-splines.


{\small
\bibliographystyle{ieee}
\bibliography{mybibfile}
}

\onecolumn
\section*{Acknowledgments}
This research is funded by the Federal Ministry of Education and Research (BMBF) of Germany (pn 02P14A035).
\\
\vspace{2em}
\vspace{-2em}
\begin{center}
\begin{large}

Supplementary Material
\end{large}
\end{center}

\begin{figure}[htb]
\centering
\includegraphics[width=1\textwidth]{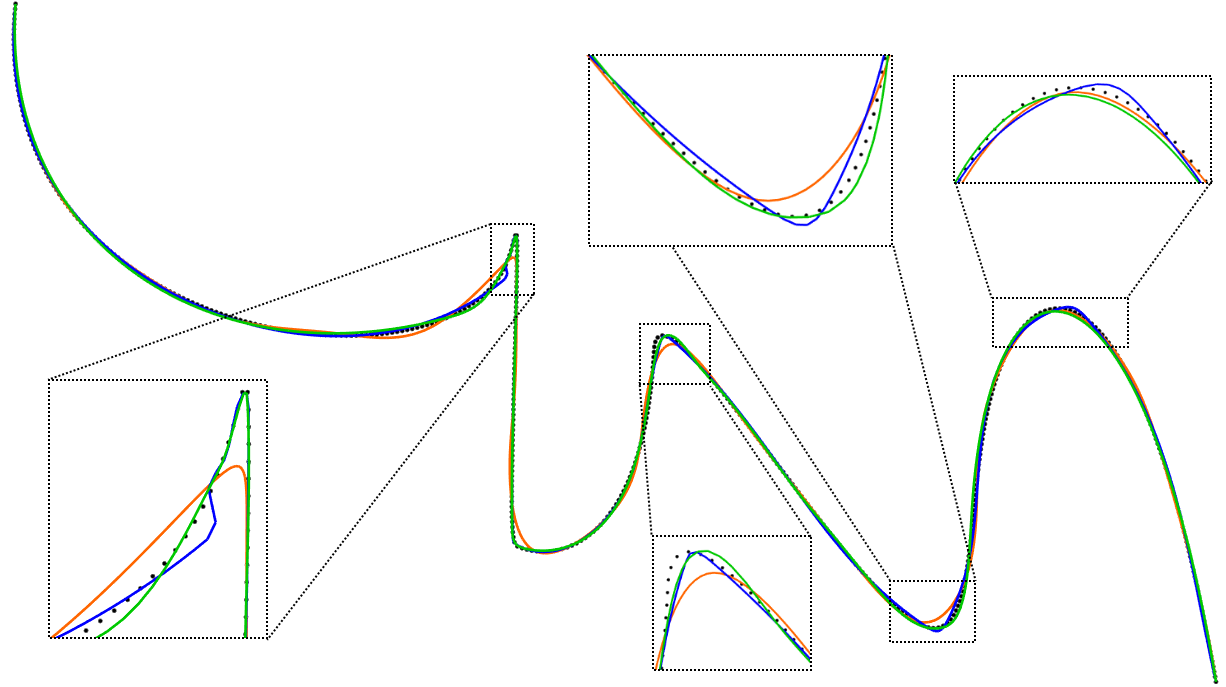}
\caption{•}
\label{fig:171}
\end{figure}

\begin{figure}[htb]
\centering
\includegraphics[width=1\textwidth]{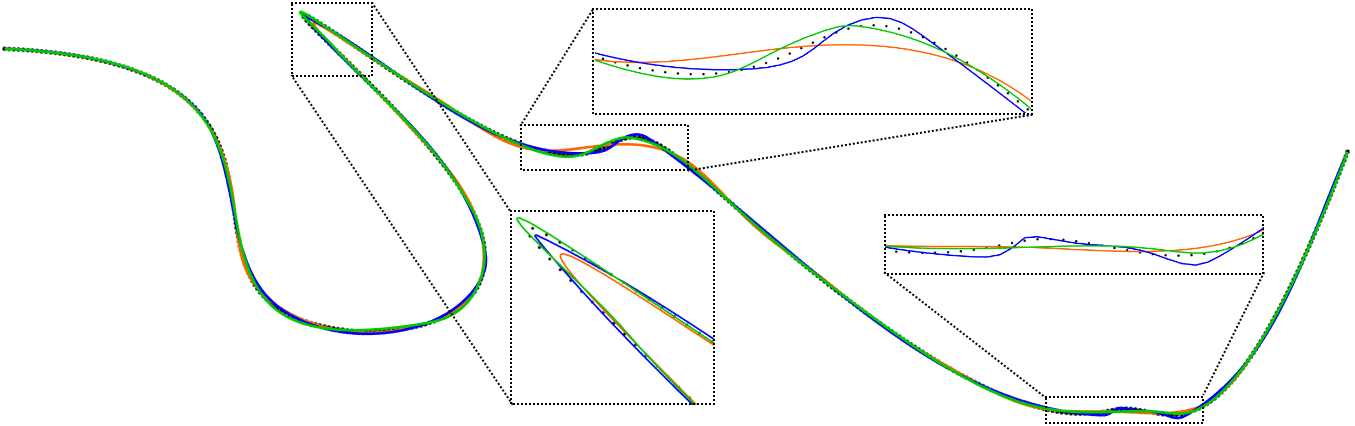}
\caption{•}
\label{fig:45}
\end{figure}

\begin{figure}[htb]
\centering
\includegraphics[width=1\textwidth]{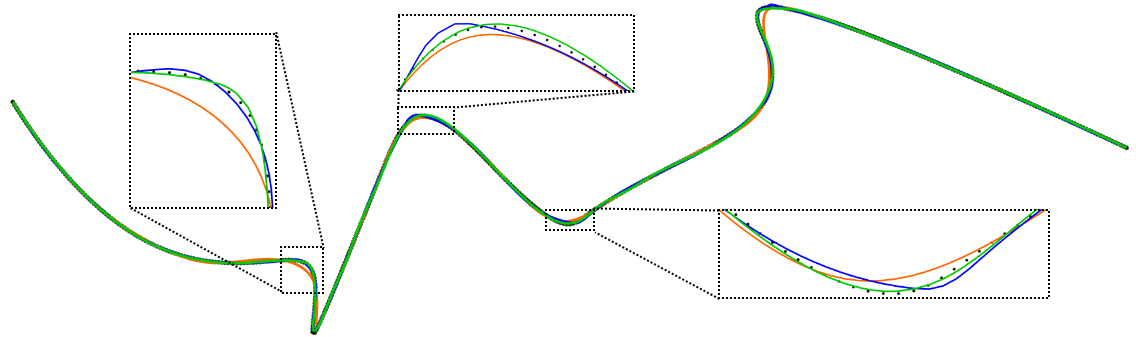}
\caption{•}
\label{fig:63}
\end{figure}

\begin{figure}[htb]
\centering
\includegraphics[width=0.7\textwidth]{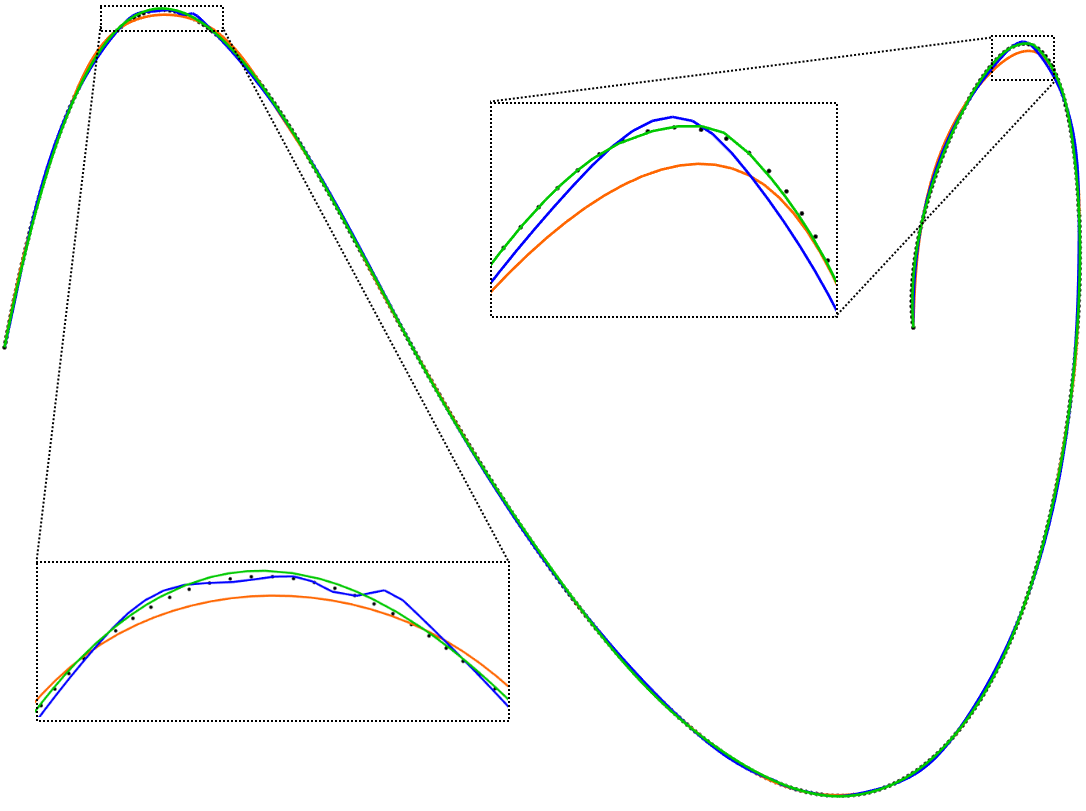}
\caption{•}
\label{fig:93}
\end{figure}

\begin{figure}[htb]
\centering
\includegraphics[width=0.7\textwidth]{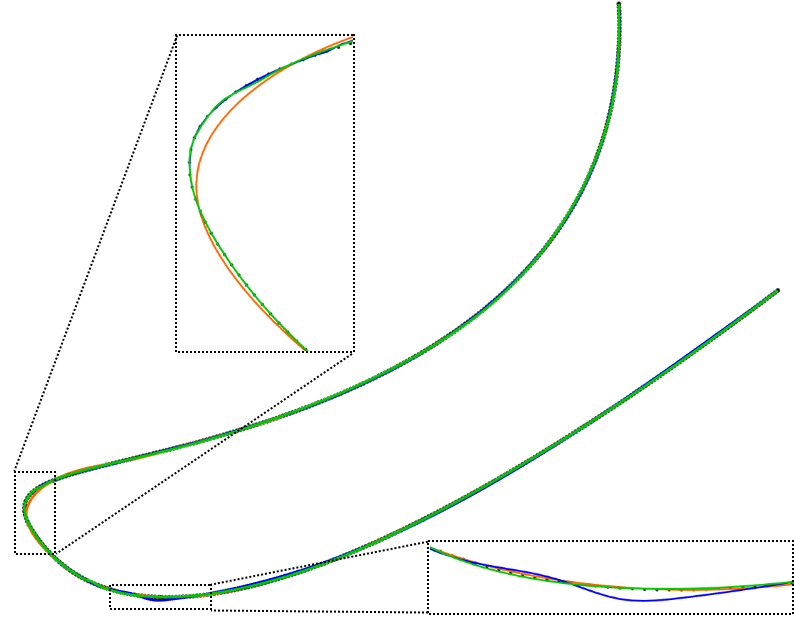}
\caption{•}
\label{fig:101}
\end{figure}

\begin{figure}[htb]
\centering
\includegraphics[width=0.9 \textwidth]{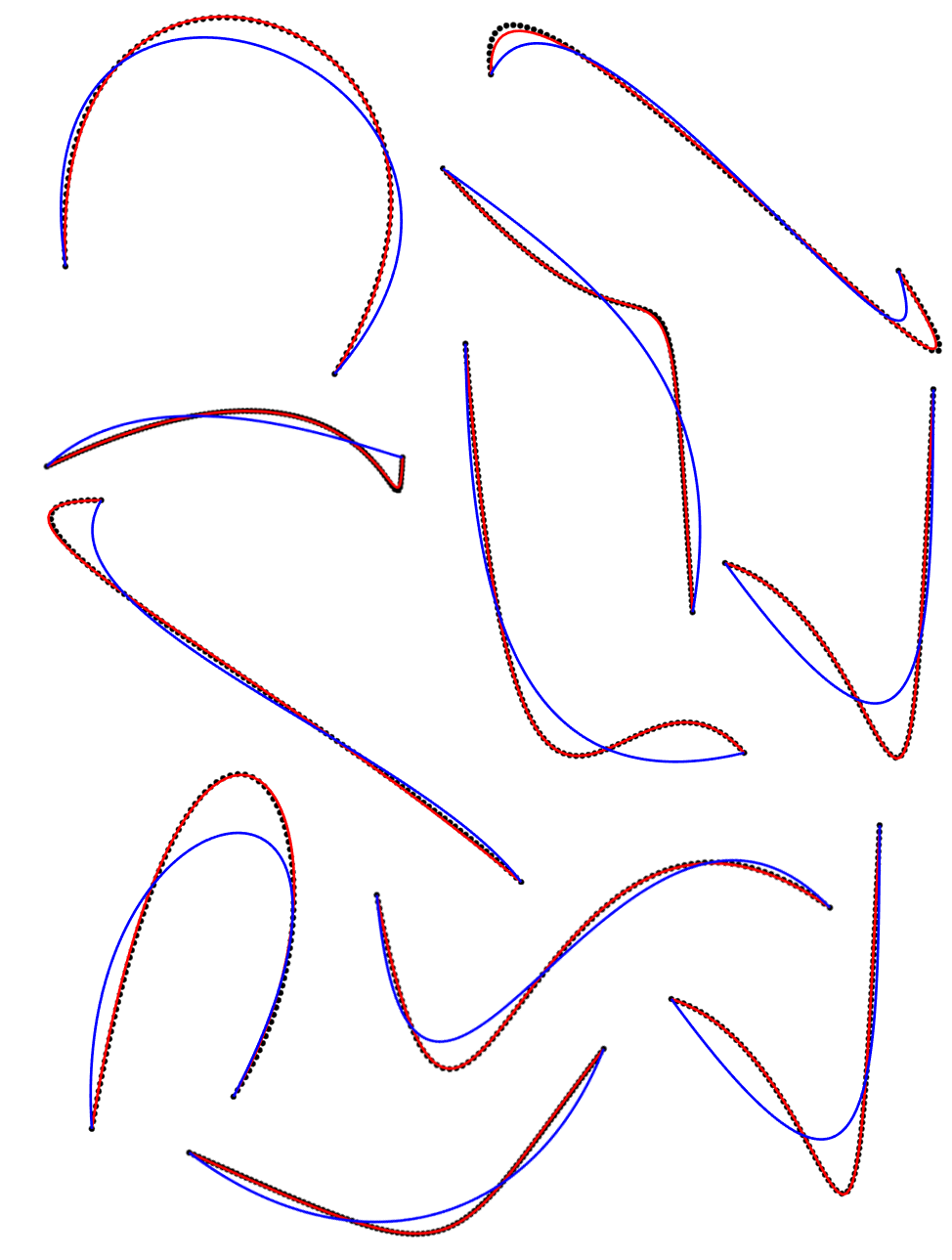}
\caption{•}
\label{fig:paramonly}
\end{figure}

\noindent
Figures \ref{fig:171}-\ref{fig:101} show additional approximation results with 23 knots for examples of \textit{evaluation set 3} by methods NKTP (orange), DPKP (blue), and PARNET (green). Black dots represent the original point sequence and boxes contain close-ups of critical regions.
In figure \ref{fig:paramonly} we present further results of parametrizations for examples
of \textit{evaluation set 1} by the centripetal method (blue) and by the PPN (red), approximated without interior knots. Again black dots represent the original point sequence.

\end{document}